\documentclass[%
 reprint,
superscriptaddress,
 amsmath,amssymb,
 aps,
prx,
]{revtex4-1}

\usepackage{graphicx}% Include figure files
\usepackage{dcolumn}% Align table columns on decimal point
\usepackage{bm}% bold math
\usepackage{color}
\usepackage{algorithm,algorithmic}
\usepackage{hyperref}% add hypertext capabilities
\newlength\myindent
\newcommand\bindent{%
  \begingroup
  \setlength{\itemindent}{\myindent}
  \addtolength{\algorithmicindent}{\myindent}
}
\newcommand\eindent{\endgroup}

\begin{document}

%\preprint{APS/123-QED}

\title{AnDi: The Anomalous Diffusion Challenge}% Force line breaks with \\

%
\author{Gorka Mu\~noz-Gil}%
\affiliation{ICFO -- Institut de Ci\`encies Fot\`oniques, The Barcelona Institute of Science and Technology, 08860 Castelldefels (Barcelona), Spain}
\author{Giovanni Volpe}
\affiliation{Department of Physics, University of Gothenburg, Gothenburg, Sweden}%
\author{Miguel Angel Garcia-March}
\affiliation{Instituto Universitario de Matem\'{a}tica Pura y Aplicada, Universitat Polit\`ecnica de Val\`encia, E-46022 Val\`encia, Spain}%
\author{Ralf Metzler}
\affiliation{Institute for Physics \& Astronomy, University of Potsdam, Karl-Liebknecht-Str 24/25, D-14476 Potsdam-Golm, Germany}%
\author{Maciej Lewenstein}
\affiliation{ICFO -- Institut de Ci\`encies Fot\`oniques, The Barcelona Institute of Science and Technology, 08860 Castelldefels (Barcelona), Spain}
\affiliation{ICREA, Lluis Companys 23, E-08010 Barcelona, Spain}
\author{Carlo Manzo}
\email{carlo.manzo@uvic.cat}
\homepage{http://www.andi-challenge.org}
\affiliation{Facultat de Ci\`encies i Tecnologia, Universitat de Vic -- Universitat Central de Catalunya (UVic-UCC), C. de la Laura,13, 08500 Vic, Spain}
\date{\today}% It is always \today, today,

\begin{abstract}
The deviation from pure Brownian motion, generally referred to as anomalous diffusion, has received large attention in the scientific literature to describe many physical scenarios. Several methods, based on classical statistics and machine learning approaches, have been developed to characterize anomalous diffusion from experimental data, which are usually acquired as particle trajectories.
With the aim to assess and compare the available methods to characterize anomalous diffusion, we have organized the Anomalous Diffusion (AnDi) Challenge (\url{http://www.andi-challenge.org/}). Specifically, the AnDi Challenge will address three different aspects of anomalous diffusion characterization, namely: (i) Inference of the anomalous diffusion exponent. (ii) Identification of the underlying diffusion model. (iii) Segmentation of trajectories. Each problem includes sub-tasks for different number of dimensions (1D, 2D and 3D). In order to compare the various methods, we have developed a dedicated open-source framework for the simulation of the anomalous diffusion trajectories that are used for the training and test datasets. The challenge was launched on March 1, 2020, and consists of three phases. Currently, the participation to the first phase is open. Submissions will be automatically evaluated and the performance of the top-scoring methods will be thoroughly analyzed and compared in an upcoming article. 

\end{abstract}


\maketitle

%\tableofcontents

\section{Introduction}

Since Albert Einstein provided a theoretical foundation~\cite{einstein1905molekularkinetischen} for Robert Brown’s observation of the movement of granules from pollen grains suspended in water~\cite{brown1828brief}, significant deviations from the laws of Brownian motion have been uncovered in a variety of animate and inanimate systems, from biology to the stock market~\cite{klafter2005anomalous,hofling2013anomalous,barkai2012single}. Anomalous diffusion, as it has come to be called, extends the concept of
Brownian motion and is connected to disordered systems, non-equilibrium phenomena, flows of energy and information, and transport in living systems~\cite{metzler2014anomalous}. Typically, anomalous diffusion is characterized by a nonlinear growth of the mean squared displacement ${\rm MSD}$ with respect to time $t$:
\begin{equation}
	{\rm MSD}(t) \sim t^\alpha.
\end{equation}
While for $\alpha = 1$ we have standard Brownian diffusion, for $\alpha < 1$ ($\alpha > 1$) we have subdiffusion (superdiffusion).
Anomalous diffusion can be generated by a variety of stochastic processes, such as: (i) the continuous-time random walk (CTRW)~\cite{scher1975anomalous}; (ii) the fractional Brownian motion (FBM)~\cite{mandelbrot1968fractional}; (iii) the L\'{e}vy walk (LW)~\cite{klafter1994levy}; (iv) the annealed transient time motion (ATTM)~\cite{massignan2014nonergodic}; (v) the scaled Brownian motion (SBM)~\cite{lim2002self}. 
Identifying the physical origin of anomalous diffusion and determining its exponent $\alpha$ are crucial to understand the nature of the systems under observation. However, the measurement of these properties from experimental trajectories is often limited especially for trajectories that are short, irregularly sampled or featuring mixed behaviors.  

\begin{figure}[b!]
    \centering
    \includegraphics[width=0.95\columnwidth]{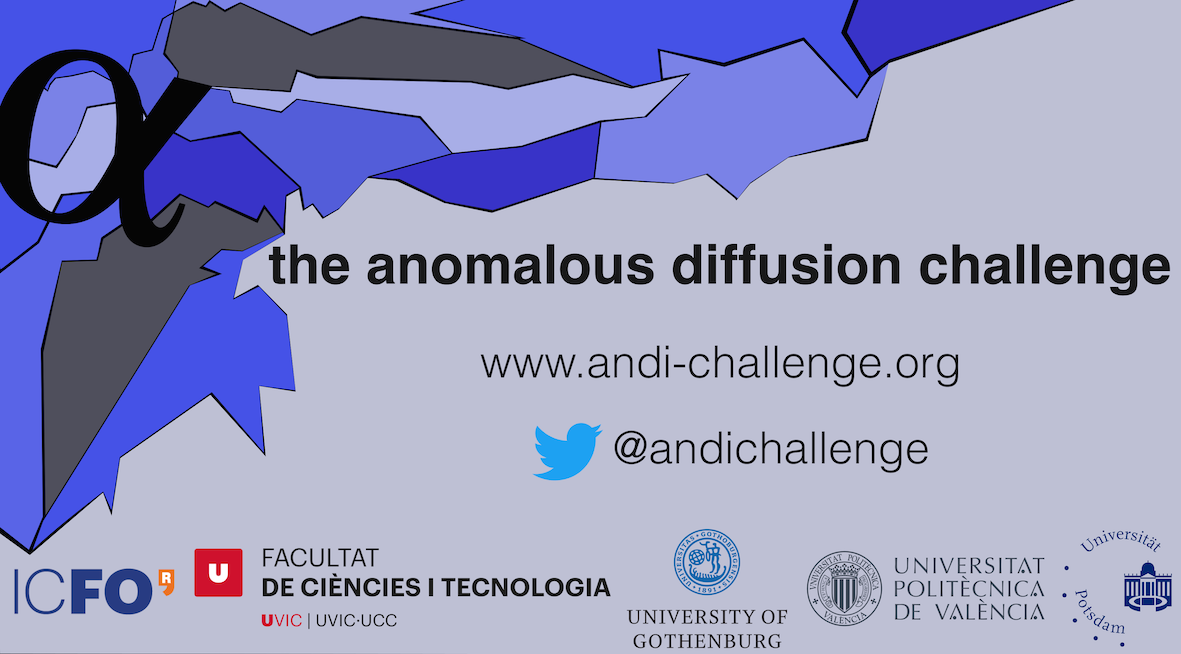}
    \caption{The Anomalous Diffusion (AnDi)  Challenge aims at bringing together a vibrating and multidisciplinary community of scientists working on the characterization of anomalous diffusion. All details can be found on \url{http://www.andi-challenge.org/}.}
    \label{fig:banner}
\end{figure}

In the last years, mainly driven by experimental advances in single-molecule  techniques and single-particle tracking~\cite{saxton1997single,manzo2015review,norregaard2017manipulation}, an ever increasing number of methods are being proposed to characterize anomalous diffusion, going beyond the classical calculation of the mean squared displacement, either through the calculation of the exponent $\alpha$ or through the determination of the underlying model. A non-exhaustive list of articles about this subject includes Refs.~\cite{granik2019single, cichos2020machine, munoz2020single, bo2019measurement, kowalek2019classification, thapa2018bayesian, krapf2018power,thapa2018bayesian, magdziarz2009fractional, burnecki2015estimating, cherstvy2019non,serov2020statistical}. 
Despite this wealth of ideas and methods, an objective comparison between them is still lacking. Therefore, we launched a joint effort in the form of a competition to benchmark these methods, to spur the invention of new methods, and ultimately to obtain new insights about anomalous diffusion.

\begin{figure}
    \centering
    \includegraphics[width=0.9\columnwidth]{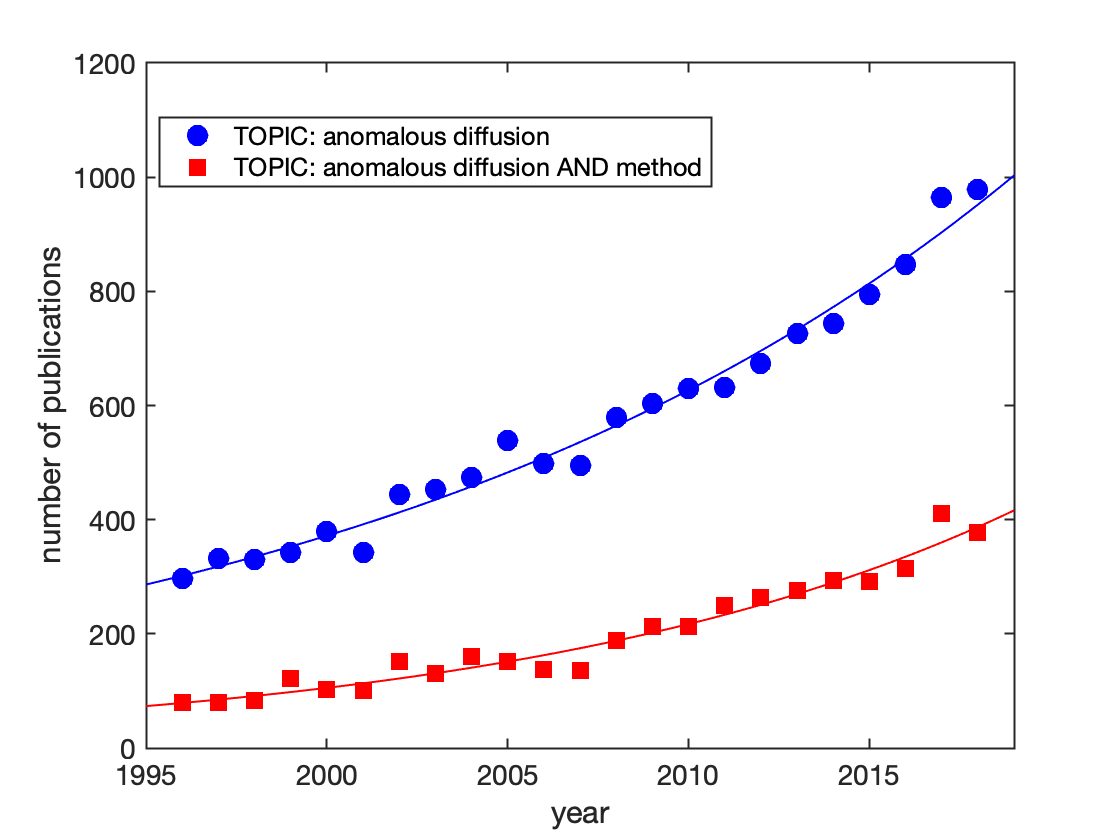}
    \caption{Number of articles per year having as a topic {\it anomalous diffusion} and {\it anomalous diffusion AND method}, as a result of a search on Web of Science. The number of articles follow an exponential growth with a doubling time of 13.3 and 9.6 years, respectively.}
    \label{fig:bibliosearch}
\end{figure}

The Anomalous Diffusion (AnDi) Challenge (\url{http://www.andi-challenge.org/}, Fig.~\ref{fig:banner}) aims at bringing together the vibrating and multidisciplinary community of scientists working on this problem (Fig.~\ref{fig:bibliosearch}). The use of the same reference datasets will allow an objective assessment of the performance of published and unpublished methods to characterize anomalous diffusion from single trajectories.

The objective of the AnDi Challenge is mainly scientific. Therefore, the top-ranking participant of each of the 9 subtasks will receive a symbolic award, consisting in the invitation to give an oral invited presentation at the AnDi workshop, which will be held at ICFO premises (Castelldefels, Spain) in February 2021. The winners will be awarded a certificate, and their travel expenses will be covered by the organization. The results of the competition will be made available to the community through a special issue of ``Journal of Physics A'' on this topic as well as through an article reviewing the insight gained during the competition.

\section{The structure of the AnDi Challenge}

The  AnDi Challenge consists of three main tasks:
\begin{description}
\item[Task 1 -- Infer anomalous diffusion exponent]
This is a regression problem, for which participants must determine the anomalous diffusion exponent $\alpha$ of each of the 10000 trajectories contained in a dataset prepared by the organizers. The dataset will include trajectories of different lengths.
\item[Task 2 -- Classify diffusion models] This task consists in a multiclass classification problem. Participants must provide five numeric values between 0 and 1, representing the scores of membership of each trajectory to the models, in the following order: ATTM; CTRW; FBM; LW; SBM. The scores must add up to 1. The dataset will include 10000 trajectories of different lengths.
\item[Task 3 -- Segment trajectories]
For this task, as in classical time-series segmentation, input trajectories showing a change of anomalous diffusion exponent $\alpha$ and/or diffusion model must be divided into two segments, in order to reveal the underlying properties of each. It is thus a combined problem, including segmentation, regression and classification. The results must include five numeric values per trajectory. The first value represents the changepoint $t_{i,{\rm calc}}$ (which is the time at which the change of behavior occurs, where 0 corresponds to the beginning of the trajectory and $T$ corresponds to the end of the trajectory). The second value must be an integer between 0 and 4, indicating the model associated to the first segment of the trajectory, according to the following convention: [0: ATTM; 1: CTRW, 2: FBM, 3: LW, 4: SBM]. The third value must correspond the value of the anomalous diffusion exponent of the first segment of the trajectory. The fourth value must be an integer between 0 and 4, indicating the model associated to the second segment of the trajectory, using the same convention used for the second value. The fifth value must be the value of the anomalous diffusion exponent of the second segment of the trajectory. If no changepoint is detected, the first value should be either 0 or $T$, and the second and the third values must be identical to the fourth and the fifth, respectively. All the trajectories will have fixed length (200 points) and at most one changepoint.
\end{description}

Each task will further include modalities for different number of dimensions (1D, 2D and 3D), for a total of 9 subtasks. Participants can submit results for an arbitrary number of subtasks.

To make the tasks more realistic, the trajectories will be corrupted with different levels of noise associated to a finite localization precision and will be scaled to have a short-time diffusion coefficient (i.e., variance of the displacement distribution) between 0 and 1. 

The challenge is hosted on the website \url{www.andi-challenge.org} that redirects to the Codalab page \url{competitions.codalab.org/competitions/23601}. Links to examples of (labeled) training dataset and sample result submissions are provided on the website. A forum is available for participants to start discussion topics related to the competition. Updates about the challenge are also provided through the official twitter account \url{@AndiChallenge}.

The competition is organized in three phases:
\begin{description}
\item[Development (March 1 to September 13, 2020)] The Development phase is intended for code development and practice only. Participants are expected to set-up their code in order to produce predictions on anomalous diffusing trajectories. A labeled dataset (``Development dataset for training'') and the code to generate trajectories are made available. Participants can submit results obtained for the ``Development dataset for Scoring'' that will be automatically evaluated and shown on the leaderboard.
\item[Validation (September 14 to October 25, 2020)] The Validation phase is meant to tweak and tune the code, so to improve its prediction capabilities of anomalous diffusion. An unlabeled ``Validation dataset'' is made available. The leaderboard will show scores for the Validation dataset only.
\item[Challenge (October 26 to November 1, 2020)] The final phase, called Challenge, has a short duration and participant can upload only a limited number of submissions corresponding to results they obtained for the ``Challenge dataset'. Only participants admitted to this phase will be included to the final ranking and publication. Results will not be scored automatically upon submission, the leaderboard will be available after the deadline.
\end{description}

Training data can be further generated by means of the code freely available at \url{https://github.com/AnDiChallenge} on GitHub. 

%%%%%%%%%%%%%%%%%%%% SIMULATED DATASET %%%%%%%%%%%%%%%%%%%%%%
\section{The simulated dataset}

Here, we will review the main features of the simulated dataset generated for the AnDi Challenge. To allow any participant to generate their own datasets, we have developed the {\tt andi-dataset} Python package, which can be installed in Python 3.6 using the package installer {\tt pip} as e.g.
\begin{verbatim}
pip install andi-datasets
\end{verbatim}
For an in-depth look into the available functions, we refer the reader to the hosting repository \url{https://github.com/AnDiChallenge/ANDI_datasets}~\cite{andigithub}.

The AnDi Challenge dataset contains trajectories generated using the 5 models considered in the challenge: CTRW, ATTM, FBM, LW, and SBM. We consider trajectories whose anomalous exponents are in the range $\alpha\in[0.05,2]$\footnote{We restricted $\alpha>0.05$ as for many models below that value the trajectories will show little to no motion.}. Note that CTRW and ATTM are subdiffusive models ($\alpha\leq 1$), LW is superdiffusive ($\alpha\geq 1$), FBM cannot have ballisitc behavior ($\alpha < 2$) and SBM covers the hole exponent range. 

In the following, we briefly present each model and its numerical implementation. While the algorithms presented below are examples for one dimensional trajectories, they are straightforwardly generalized to two and three dimensions. In such cases, care has to be taken at sampling every direction uniformly, as all methods considered are isotropic. This means that for two dimensions, the steps have to be sampled uniformly on a circle of radius $x_i$, where $x_i$ is the length of the step, and for three dimensions steps on a sphere of radius $x_i$.

%%%%%%%%% CTRW %%%%%%%%%%%
\subsection{Continuous time random walk}
The continuous time random walk (CTRW) models the motion of a particle for which the time between subsequent steps, defined as the \textit{waiting time}, is irregular~\cite{scher1975anomalous}. More precisely, we consider a CTRW in which the waiting times follow the distribution $\psi(t)\sim t^{-\sigma}$ and the step lengths are sampled from a Gaussian distribution with variance $\sqrt{D}$ and mean zero. In such case, the anomalous exponent is $\alpha = \sigma-1$. The algorithm used to simulate CTRW trajectories is presented in Algorithm~\ref{alg:ctrw}.

\begin{algorithm}[H]
\caption{Generate CTRW trajectory}
\label{alg:ctrw}
\begin{algorithmic}
\STATE{\bfseries Input:}
\STATE{$\qquad\quad$length of the trajectory $T$}
\STATE{$\qquad\quad$anomalous exponent $\alpha$}
\STATE{$\qquad\quad$diffusion coefficient $D$}
\STATE{\bfseries Define:}
\STATE{$\qquad\quad \vec{x} \rightarrow$ empty vector}
\STATE{$\qquad\quad \vec{t} \rightarrow$ empty vector}
\STATE{$\qquad\quad N(\sigma, \mu) \rightarrow$ Gaussian random number generator with variance $\sigma$ and mean $\mu$}
\STATE{$i = 0$}
\WHILE {$\tau<T$}
\STATE $t_i \leftarrow$  sample randomly from  $\psi(t)=t^{-\sigma}$
\STATE $x_i \leftarrow x_{i-1} + N(1,0)$
\STATE $\tau \leftarrow \tau + t_i$
\STATE $i \leftarrow i+1$
\ENDWHILE
\STATE{\bfseries Return:} $\vec{x}, \ \vec{t}$
\end{algorithmic}
\end{algorithm}

%%%%%%%%% ATTM %%%%%%%%%%%
\subsection{Annealed transient time}
The annealed transient time (ATTM) considers the motion of a Brownian particle whose diffusion coefficient varies either in time or space~\cite{massignan2014nonergodic}. For the AnDi dataset, we considered the \textit{regime I} of such a model (see Ref.~\cite{massignan2014nonergodic}), which considers the time variation. This means considering the walk of a particle which diffuses for a time $\tau_1$ with diffusion coefficient $D_1$, then for $\tau_2$ with $D_2$, etc. The diffusion coefficients are sampled from the distribution $P(D)\sim D^{\sigma-1}$. If one considers that $\tau_i = D_i^{-\gamma}$, the anomalous exponent is shown to be $\alpha=\sigma/\gamma$. The numerical implementation of the ATTM model is given in Algorithm~\ref{alg:attm}.

\begin{algorithm}[H]
\caption{Generate ATTM trajectory}
\begin{algorithmic}
\label{alg:attm}
\STATE{\bfseries Input:}
\STATE{$\qquad$length of the trajectory $T$}
\STATE{$\qquad$anomalous exponent $\alpha$}
\STATE{$\qquad$sampling time $\Delta t$}
\STATE{\bfseries Define:}
\bindent
\STATE{$\sigma, \gamma \rightarrow$ generate randomly s.t.  $\alpha = \sigma/\gamma$ and  $\sigma < \gamma < \sigma+1$}
\STATE{$\mbox{BM}(D,t,\Delta t) \rightarrow$ generates a Brownian motion trajectory of length $t$ with diffusion coefficient $D$, sampled at time intervals $\Delta t$}
\STATE{$\vec{x} \rightarrow$ empty vector}
\eindent
\WHILE {$t<T$}
\STATE $D \leftarrow$ sample randomly from  $P(D)=D^{\sigma-1}$
\STATE $\tau \leftarrow D^{-\gamma}$
\STATE $x_i,...,x_{i+\tau/\Delta t} \leftarrow \mbox{BM}(D,\tau,\Delta t)$
\STATE $i \leftarrow i+\tau+1$
\ENDWHILE
\STATE{\bfseries Return:} $\vec{x}$
\end{algorithmic}
\end{algorithm}

%%%%%%%%%%% FBM %%%%%%%%%%%
\subsection{Fractional Brownian Motion}
Fractional Brownian motion (FBM) arises from  the Langevin equation $\frac{dx(t)}{dt}=\xi_{fGn}(t)$, where $\xi_{fGn}$ is a fractional Gaussian noise~\cite{mandelbrot1968fractional}. The latter has a standard normal distribution but has power-law correlations:
\begin{equation}
 \left< \xi_{fGn}(t_1)\xi_{fGn}(t_2)\right> = \alpha (\alpha-1)K_\alpha|t_1-t_2|^{\alpha-2},  
\end{equation}
for $t_1,t_2>0$ and $t_1\neq t_2$, where $\alpha$ is, again, the anomalous exponent of the generated trajectory. Fractional Brownian Motion has two regimes: one  where the noise is persistent and positively correlated ($1<\alpha<2$) and one where the is noise is antipersistent and negatively correlated ($0<\alpha<1$). For $\alpha = 1$ the noise is uncorrelated, hence the FBM converges to usual Brownian motion.

There exist currently various numerical approaches to exactly solve the FBM Langevin equation. We use in this case the Davies-Harte method~\cite{davies1987} and the Hosking method~\cite{hosking1984modeling} via the {\tt FBM} Python package~\cite{fbmweb}. For details on the numerical implementations, we refer the reader to the previous references.

%%%%%%%%%%% LW %%%%%%%%%%%
\subsection{Lévy Walks}
The Lévy walk (LW) shares some similarities with the CTRW, in the sense that the time between steps is also irregular~\cite{klafter1994levy}. However, the distribution of displacements for a LW is not Gaussian, different from the case for a CTRW. We will consider here the case in which the \textit{flight times}, i.e. the time between steps, are retrieved from the distribution $\psi(t)\sim t^{-\sigma-1}$. The displacements, or \textit{step lenghts}, are correlated with the flight times s.t. $\Psi(\delta x, t) = \frac{1}{2}\psi(t)\delta(|\delta x|-vt)$, where $v$ is the velocity. From here, one can show that the anomalous exponent is given by
\begin{equation}
\alpha =
  \begin{cases}
  2       & \mbox{if} \ 0<\sigma<1 \\
  3-\sigma  & \mbox{if} \ 1<\sigma<2.
  \end{cases}
\end{equation}
The details of the numerical implementation for the LW are given in Algorithm~\ref{alg:lw}.

\begin{algorithm}[H]
\caption{Generate LW trajectory}
\label{alg:lw}
\begin{algorithmic}
\STATE{\bfseries Input:}
\bindent
\STATE{length of the trajectory $T$}
\STATE{anomalous exponent $\alpha$}
\eindent
\STATE{\bfseries Define:}
\bindent
\STATE{$\vec{x} \leftarrow$ empty vector}
\STATE{$\vec{t} \leftarrow$ empty vector}
\STATE{$v \leftarrow$ random number $\in (0,10]$}
\eindent
\STATE{$i = 0$}
\WHILE {$\tau<T$}
\STATE $t_i \leftarrow$ sample from $\psi(t)\sim t^{-\sigma-1}$
\STATE $x_i \leftarrow (-1)^rvt$, where $r = \{0,1\}$
\STATE $t \leftarrow \tau + t_i$
\STATE $i \leftarrow i+1$
\ENDWHILE
\STATE{\bfseries Return:} $\vec{x},  \vec{t}$
\end{algorithmic}
\end{algorithm}

%%%%%%%%%%% SBM %%%%%%%%%%%%%%%%%%%%
\subsection{Scaled Brownian Motion}
Scaled Brownian motion (SBM) is a process that shows a time-dependent diffusivity $K(t)$, with the following Langevin equation
\begin{equation}
\frac{dx(t)}{dt}=\sqrt{2K(t)}\xi(t),
\end{equation}
where $\xi(t)$ is white Gaussian noise~\cite{lim2002self}. For the case in which $K(t)$ has a power-law dependence with $t$,  $K(t)=\alpha K_\alpha t^{\alpha-1}$, the ensemble MSD follows $\left< x^2(t) \right>_N\approx K_\alpha t^\alpha$ with $K_\alpha = \Gamma(1+\alpha)K_\alpha$. This means that the anomalous exponent is $\alpha$. The numerical implementation of SBM is presented in Algorithm~\ref{alg:sbm}.

\begin{algorithm}[H]
\caption{Generate SBM trajectory}
\label{alg:sbm}
\begin{algorithmic}
\STATE{\bfseries Input:}
\bindent
\STATE{length of the trajectory $T$}
\STATE{anomalous exponent $\alpha$}
\eindent
\STATE{\bfseries Define:}
\bindent
\STATE{{\tt erfcinv}$(\vec{a}) \leftarrow$ Inverse complementary {\rm erf} of $\vec{a}$ }
\STATE{\textit{U}$(L) \leftarrow$ returns $L$ uniform random numbers $\in [0,1]$}
\eindent
\STATE{\bfseries Calculate:}
\bindent
\STATE{$\overrightarrow{\Delta x} \leftarrow (0,1,...,T)^\alpha - (1,...,T+1)^\alpha$}
\STATE{$\overrightarrow{\Delta x} \leftarrow 2\sqrt{2} U(L) \overrightarrow{\Delta x}$},
\STATE{$\vec{x} \leftarrow \mbox{{\tt cumsum}}(\overrightarrow{\Delta x})$}.
\eindent
\STATE{\bfseries Return:} $\vec{x}$
\end{algorithmic}
\end{algorithm}

%%%%%%%%%%%%% EVALUATION OF THE PERFORMANCE %%%%%%%%%%%%%%%%%%%
\section{\label{sec:evaluation} Evaluation of the performance}
The ranking of the submissions will be performed using standard metrics, as described below for each task. 

\subsection{Task 1 -- Infer anomalous diffusion exponent}
The results will be evaluated by the calculation of the mean absolute error (${\rm MAE}$)
\begin{equation}
{\rm MAE} = \frac{1}{N} \sum_{i=1}^{N}{| \alpha_{i,\mbox{calc}} - \alpha_{i,\mbox{GT}} |},
\end{equation}
where $N$ is the number of trajectories in the dataset, $\alpha_{i,\mbox{calc}}$ and $\alpha_{i,\mbox{GT}}$ represent the calculated and ground truth values of the anomalous exponent of the $i$-th trajectory, respectively.

\subsection{Task 2 -- Classify diffusion models}
The results will be evaluated by the calculation of the $F_1$ score, i.e. the harmonic mean of the precision and the recall:
\begin{equation}
F_1 = 2 \cdot \frac{{\rm precision} \cdot {\rm recall}}{{\rm precision} + {\rm recall}},
\end{equation}
where ${\rm precision} = \frac{{\rm TP}}{{\rm TP}+{\rm FP}}$ and ${\rm recall}=\frac{{\rm TP}}{{\rm TP}+{\rm FN}}$, with ${\rm TP}$, ${\rm FP}$ and ${\rm FN}$ being the true positive, false positive  and false negative rates, respectively. In particular, the ``micro'' version of the $F_1$ score provided in the Scikit-learn Python's library will be used and the metrics will be calculated globally by counting the total ${\rm TP}$, ${\rm FP}$ and ${\rm FN}$.

\subsection{Task 3 -- Segment trajectories}
For this task, in addition to the ${\rm MAE}$ and $F_1$ scores, we will also calculate the root mean squared error (${\rm RMSE}$) of the changepoint localization:
\begin{equation} 
{\rm RMSE} = \frac{1}{N} \sqrt{\sum_{i=1}^{N} \Big(t_{i,{\rm calc}} - t_{i,{\rm GT}} \Big)^2},
\end{equation}
where $t_{i,{\rm calc}}$ and $t_{i,{\rm GT}}$ represent the calculated and ground truth values of the changepoint relative position, respectively.
For the ranking, the precision in determining the changepoint position, the anomalous diffusion exponent $\alpha$, and the diffusion model will be summarized in a unique metric given by the mean reciprocal rank: 
\begin{equation}
{\rm MRR} = \frac{1}{3}\cdot 
\left(\frac{1}{{\rm rank}_{\rm MAE} } +\frac{1}{{\rm rank}_{F_1}}+\frac{1}{{\rm rank}_{\rm RMSE}}\right).
\end{equation}

\subsection{Leaderboard}
The Codalab leaderboard will show the average rank. Participants can submit results for an arbitrary number of subtasks. In the leaderboard, submission lacking results for some task/subtask will be scored by default with ${\rm MAE}=100$ (tasks 1 and 3), $F_1=0$ (tasks 2 and 3), and ${\rm RMSE}=200$ (task 3).

Even though these metrics are used for the official ranking, the performance of the methods might be further analyzed in other forms for research purposes, e.g., at varying trajectory parameters, such as length or noise, or by using other metrics (e.g. ROC-AUC, top-(n) accuracy, $F_{\beta}$).

\section{Dissemination and awards}
The top participants will be invited to contribute to a joint article describing and summarizing the methods used and the results obtained for the challenge. The paper will be submitted to the arXiv first and then to an indexed journal. The paper will have sufficient detail to ensure the reproducibility of the presented methods. 
In order to be included as authors, participants must provide a detailed description of their methods and be available to clarify any doubt that might arise concerning the methods, the code, and/or the results. 
In addition, they must provide an open-source version of the code used to analyze the ``AnDi Challenge dataset'' under an OSI-approved license such as, for instance, Apache 2.0, MIT or BSD-like license.
All participants will be invited to contribute with articles to a special issue in ``Journal of Physics A'' to be announced soon.
All participants will also be invited to attend the AnDi workshop to be held at ICFO premises (Castelldefels, Spain) in February 2021.

The top-ranking participants of each of the 9 award-winning subtasks may qualify for awards (travel award, invited oral contribution and award certificate). To receive the award, they will need to provide a link to a repository containing: (i) An open-source version of the code used to analyze the test set under an OSI-approved license such as, for instance, Apache 2.0, MIT or BSD-like license.
(ii) A brief description of the algorithm and instructions to run the code.
The winners of each subtask will be determined according to a ranking calculated as described in Sec.~\ref{sec:evaluation}. In the case of a tie, the prize will go to the participants who submitted their entry first.
The travel awards will cover the expenses to attend the workshop organized in conjunction with the challenge, including airfare (economy class), hotel and workshop registration. The award is conditioned on attending the workshop and giving an invited oral presentation of the methods used in the challenge. If the winner is a team, the award will cover the expenses of only one of the team members.

\section{Code and Data Availability}
The code and the data provided for the AnDi challenge on the Codalab website and at AnDiChallenge on GitHub are licensed under Attribution-NonCommercial-ShareAlike 4.0 International. The data can be used for scientific and educational purposes. Any commercial use of data is forbidden. Appropriate citations must be included in scientific publications (journal publications, conference papers, technical reports, presentations at conferences and meetings, thesis, etc.) that use the code and/or the data shared in this challenge. The citation must refer to this article and to the doi: 10.5281/zenodo.3707702 ~\cite{andigithub}, and later also to other publications describing the results of the AnDi challenge. Participants are encouraged notify the organizers of the AnDi challenge about any publication that is even partly based on the results or data published on this site in order to maintain a list of publications associated with the challenge.

\begin{acknowledgments}
We thank S. Thapa for sharing his implementation of the SBM model.
\end{acknowledgments}

%\appendix
%\section{Appendixes}

%\nocite{*}

\bibliography{biblio}% Produces the bibliography via BibTeX.

\end{document}